\documentclass[11pt,twoside]{article}
\usepackage{graphicx}
\usepackage{amsmath}
\usepackage{amssymb}
\usepackage{cite}

\begin{document}
\newcommand{\pst}{\hspace*{1.5em}}

\newcommand{\rigmark}{\em Journal of Russian Laser Research}
\newcommand{\lemark}{\em Volume 30, Number 5, 2009}

\newcommand{\be}{\begin{equation}}
\newcommand{\ee}{\end{equation}}
\newcommand{\bm}{\boldmath}
\newcommand{\ds}{\displaystyle}
\newcommand{\bea}{\begin{eqnarray}}
\newcommand{\eea}{\end{eqnarray}}
\newcommand{\ba}{\begin{array}}
\newcommand{\ea}{\end{array}}
\newcommand{\arcsinh}{\mathop{\rm arcsinh}\nolimits}
\newcommand{\arctanh}{\mathop{\rm arctanh}\nolimits}
\newcommand{\bc}{\begin{center}}
\newcommand{\ec}{\end{center}}

\thispagestyle{plain}

\label{sh}


\begin{center} {\Large \bf
\begin{tabular}{c}
GENERALIZED ARAGO-FRESNEL LAWS:
\\[-1mm]
THE EME-FLOW-LINE DESCRIPTION
\end{tabular}
 } \end{center}

\bigskip

\bigskip

\begin{center} {\bf
M. Bo\v zi\'c$^1$, M. Davidovi\'c$^2$, T. L. Dimitrova$^{3,4}$,
S. Miret-Art\'es$^5$,\\[+1mm]
A. S. Sanz$^5$ and A. Weis$^4$
}\end{center}

\medskip

\begin{center}
{\it
$^1$Institute of Physics, University of Belgrade, Pregrevica 118,
11080 Belgrade, Serbia

\smallskip

$^2$Faculty of Civil Engineering, University of Belgrade,\\
Bulevar Kralja Aleksandra 73, Belgrade, Serbia

\smallskip

$^3$Department of Experimental Physics, Plovdiv University ``Paisii
Hilendarski'',\\ Tsar Assen Str. 24 Plovdiv, Bulgaria

\smallskip

$^4$Physics Department, University of Fribourg,\\ Chemin du Mus\'ee 3,
CH-1700 Fribourg, Switzerland

\smallskip

$^5$Instituto de F\'{\i}sica Fundamental, Consejo Superior de
Investigaciones Cient\'{\i}ficas,\\ Serrano 123, 28006 - Madrid, Spain

}
\smallskip

e-mails: bozic@ipb.ac.rs, milena@grf.bg.ac.rs,
doradimitrova@uni-plovdiv.bg,\\ s.miret@imaff.cfmac.csic.es,
asanz@imaff.cfmac.csic.es, antoine.weis@unifr.ch\\
\end{center}

\begin{abstract}\noindent
We study experimentally and theoretically the influence of light
polarization on the interference patterns behind a diffracting
grating.
Different states of polarization and configurations are been
considered.
The experiments are analyzed in terms of electromagnetic energy (EME)
flow lines, which can be eventually identified with the paths followed
by photons.
This gives rise to a novel trajectory interpretation of the
Arago-Fresnel laws for polarized light, which we compare with
interpretations based on the concept of ``which-way'' (or
``which-slit'') information.
\end{abstract}

\medskip

\noindent{\bf Keywords:} Arago-Fresnel laws, which-way information,
electromagnetic-energy-flow line, photon interference, quantum optics


\section{Introduction}
\label{sec1}

Experiments with polarized light are well known since the beginning of
the 19th century. By this time, Arago and Fresnel \cite{arago} found a
series of results related to Young's interference experiment when the
latter is carried out with polarized light.
These results were summarized in the form of four laws, known as the
Arago-Fresnel laws \cite{arago,henry,barakat,mujat}.
According to these laws, two beams of the same linear polarization
interfere with each other just as natural rays do.
However, no interference pattern is observable if the two interfering
beams are linearly polarized in orthogonal directions. In the 1970s,
a series of experiments with linearly and elliptically polarized laser
light were performed, illustrating the Arago-Fresnel laws
\cite{hunt,pescetti}.
Actually, the original Arago-Fresnel laws were generalized to include
elliptically polarized light: two beams with the same polarization
state interfere with each other just as natural rays do, but no
interference pattern will be observable if the two interfering beams
are elliptically polarized, with opposite handedness and mutually
orthogonal major axes.

The standard interpretation given to the disappearance of interference
after inserting mutually orthogonal polarizers after the slits is
usually based on the Copenhagen notion of the external observer's
knowledge (information) about the photon paths, i.e., the slit
traversed by the photon in its way to the detection screen.
If, after allowing the two diffracted beams to reach the screen,
additional polarizers are inserted between the slits and the screen,
interference emerges again \cite{hunt,kanseri}.
This reappearance is attributed to the erasure of the observer's
which-way information, in this case actually ``which-slit'' information
\cite{scully,walborn}.

Recently, lecture demonstrations of interference \cite{dimitrova1} and
quantum erasure \cite{dimitrova2,dimitrova3} with single photons have
been experimentally realized by Dimitrova and Weis.
In both cases, experiments were first carried out with strong light.
Then, in a second step, the same experiments were performed with
strongly attenuated light by means of single photon counting.
In this way the formation of the corresponding pattern by the
progressive detection of single photons was observed, as it has
also be done with matter particles \cite{tonomura,shimuzu}.
Motivated by the feasibility of this kind of experiments, Sanz {\it et
al.}\ \cite{sanz1} have recently proposed a description based on
electromagnetic energy (EME) flow lines to analyze the formation or
disappearance of interference features depending on polarization.
As shown, within this theoretical framework it is possible to
understand experiments like the one just mentioned on the basis of the
topology displayed by the EME flow lines in their way from the grating
to the screen and on how the presence and arrangement of polarizers
behind the grating influences them.
This approach, suggested as a tool to analyze and explore basic
experiments in quantum optics, thus combines the ideas and methods from
the hydrodynamical formulation of Maxwell's equations \cite{bialynicki}
and Schr\"odinger's wave mechanics.
A similar synthesis has also been exploited by M.\ Man'ko and coworkers
to study charged particle beam optics \cite{fedele} and to describe the
classical propagation of electromagnetic waves through optical fibers
\cite{manko}.

Earlier, EME flow lines for a particular polarization of light were
determined by Braunbek and Laukien \cite{braunbek,born-wolf} in the
case of a half plate or, later, following a similar approach, by
Prosser \cite{prosser1a,prosser1b,prosser2} for the case of a single
and double-slit interference.
More recently, some of us \cite{davidovic} considered the case of
linearly polarized light and provided a method to systematically
compute the evolution of the EME flow lines by means of the transverse
momentum approach \cite{arsenovic}.
Also, Gondran and Gondran \cite{gondran} have considered the
EME-flow-line approach to study diffraction by a circular aperture
and a circular opaque disk, the latter giving rise to the well-known
Poisson-Arago spot phenomenon.
Authors put this result in the historical perspective by showing that
EME flow lines provide a complementary answer to Frenel's answer to
the questions presented in 1818 by the French Academy ``deduce by
mathematical induction the movements of the rays during their crossing
near the bodies'' \cite{fresnel}.

The four Arago-Fresnel laws governing the interference of polarized
light were determined experimentally using natural and linearly
polarized light.
As mentioned above, experimental verifications of these laws were done
with intense laser light.
Here, we analyze these laws under the conditions of single-photon count
experiments, providing a description and explanation in terms of EME
flow lines.
Accordingly, the organization of this work which encompasses both
experiment and theory is as follows.
Thus, the experimental setup and findings are described in
Sec.~\ref{sec2}.
In Sec.~\ref{sec3}, in order to be self-contained, we present a
brief overview of the EME-flow-line approach, described in detail
in \cite{sanz1}.
In particular, we focus on how the presence of polarizes influences
both electromagnetic (EM) field and EME flow lines.
In Sect.~\ref{sec4} we show the EME flow lines obtained with the
conditions used in the experiments.
Finally, in Sec.~\ref{sec5} we provide a discussion on the
interpretation of the experimental results in terms of EME flow lines
as well as a comparison between this interpretation and that one based
on the concepts of ``which-slit'' information and quantum erasure
\cite{scully,walborn}.


\section{Influence of the polarization state on the double-slit
interference pattern: Experimental facts}
\label{sec2}

\begin{figure}
 \bc
 \includegraphics[width=14cm]{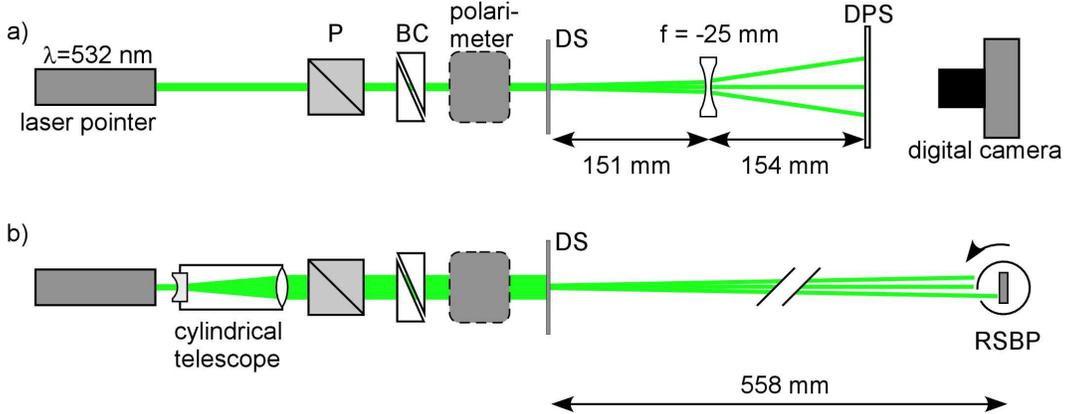}
 \ec
 \vspace{-4mm}
 \caption{\label{fig1} Schemes of the experimental setups used.
  Abbreviations: P: linear polarizer, DS: double-slit, BC: Babinet
  compensator, DPS: diffusive projection screen, RSBP: rotating slit
  beam profiler.}
\end{figure}

\begin{figure}
 \bc
 \includegraphics[height=19cm]{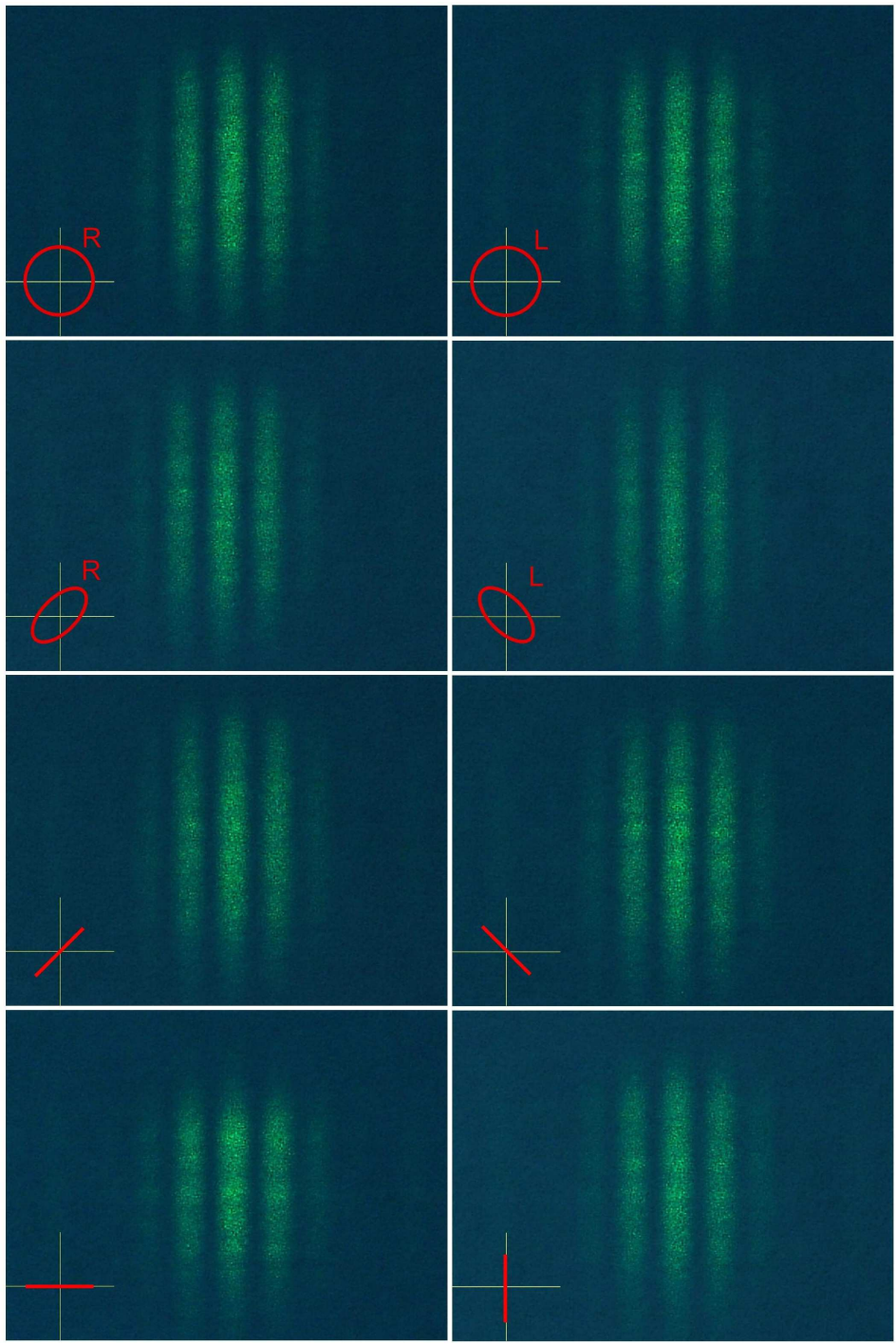}
 \ec
 \vspace{-4mm}
 \caption{\label{fig2} Double-slit interference patterns for eight
  polarization states of the incident laser light, where insets show
  right and left circular, right and left elliptic, and linear with
  angles $\alpha_p = 45^\circ, 135^\circ, 180^\circ$, and $90^\circ$.}
\end{figure}

Experiments were performed at the University of Fribourg.
Light from a green laser module ($\lambda = 532.5$~nm) was sent through
a polarizing system consisting of a linear polarizer and a Babinet
compensator (BC), which allowed the realization of arbitrary linear
and elliptical polarizations (Figs.~\ref{fig1}a and b).
The beam polarization was measured with a repositionable commercial
polarimeter that was removed for the actual diffraction measurements.
The double-slit consisted of two 18~mm long slits of 100~$\mu$m, whose
centers were separated by 250~$\mu$m.
For the experiments shown in Fig.~\ref{fig2}, the laser beam has a
nearly circular Gaussian intensity profile,
\be
 I(r) = I_0 e^{-2r^2/w^2} ,
 \label{eq1}
\ee
with $w = 1.4$~mm.

Two-dimensional interference patterns were recorded with the setup of
Fig.~\ref{fig1}a for various incident light polarizations.
The diffraction patterns were projected onto a thin sheet of paper and
their intensity distribution photographed by a digital camera.
The results are shown in Fig.~\ref{fig2} for eight circular, elliptical
and linear polarization states.
In a second experiment, the interference pattern produced by
the expanded beam was recorded by a scanning slit beam profiler
(Fig.~\ref{fig1}b) for eight polarization states.
The signal of the beam profiler corresponds to the $x$-dependence of
the intensity integrated over the $z$-direction.

\begin{figure}
 \bc
 \includegraphics[width=16cm]{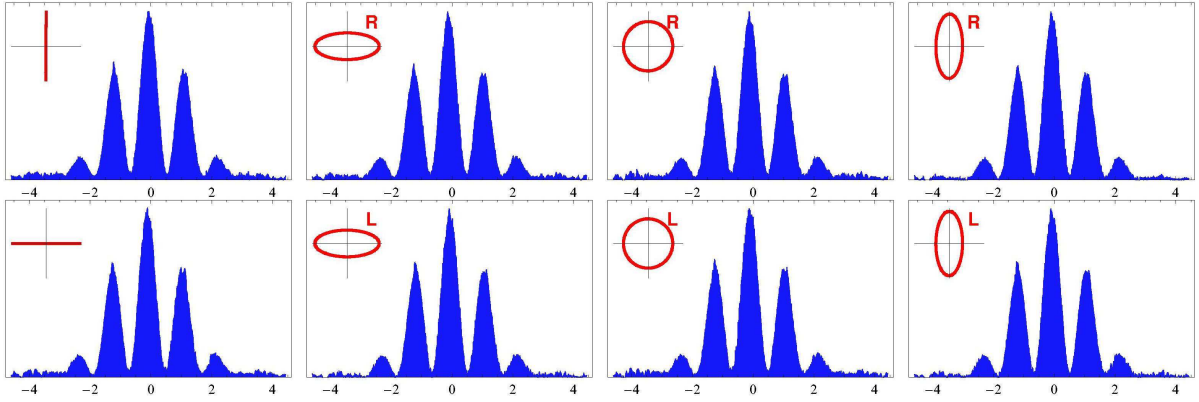}
 \ec
 \vspace{-4mm}
 \caption{\label{fig3} Interference patterns from a double-slit (0.1~mm
  slit width, separated by 0.25~mm) along the $x$-direction [in mm]
  obtained from the intensity (EME density) integrated over the
  $z$-direction for eight polarization states (shown as insets) of
  the incident light.}
\end{figure}

\begin{figure}
 \bc
 \includegraphics[height=6cm]{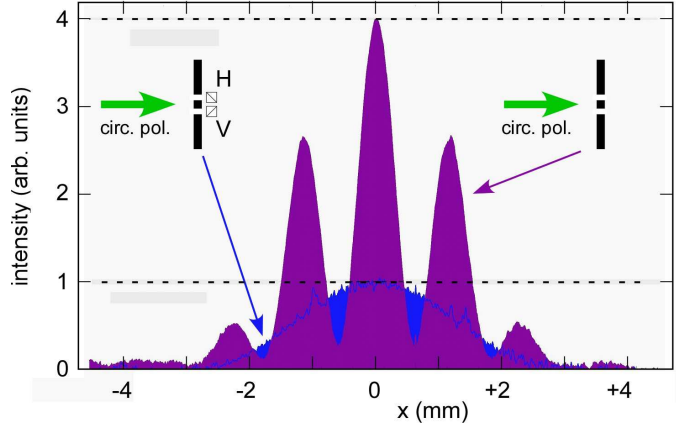}
 \ec
 \vspace{-4mm}
 \caption{\label{fig4} Intensity (EME density) as a function of $x$ at
  the screen behind the double-slit (0.1~mm slit width, separated by
  0.25~mm) illuminated with circularly polarized light in two cases:
  with H and V polarizers (blue curve) just behind each slit and
  without them (purple curve).}
\end{figure}

The results of Fig.~\ref{fig2} seem to indicate a slight polarization
dependent transversal shift.
This could be traced back to a systematic beam displacement during
adjustments of the Babinet compensator.
The results of Fig.~\ref{fig3} were obtained after resolving this
problem.
Fig.~\ref{fig2} ---and more convincingly Fig.~\ref{fig3}--- show that
the interference pattern does not depend on the state of polarization
of the incident laser light.
In a third experiment ---carried out with incident circularly polarized
light---, we have inserted a horizontal and a vertical polarizer after
each slit, respectively.
Figure~\ref{fig4} shows the diffraction patterns obtained with and
without these polarizers.
The peak intensities are in the ratio 4:1, as expected.
We note that single photon interference experiments with
mutually orthogonally polarized waves are discussed in
\cite{dimitrova2,dimitrova3} in the frame of quantum erasure processes.

\section{Theoretical background}
\label{sec3}


\subsection{EM field and EME density behind a double-slit grating for
different incident polarizations}
\label{sec31}

Consider the EM field behind a grating situated on the $XZ$-plane, at
$y = 0$, with slits parallel to the $z$-axis, whose width ($\delta$)
is much larger along the $z$-direction than along the $x$-direction.
Accordingly, we can assume the EME density to be independent of the
$z$-coordinate and, therefore, the electric and magnetic fields will
not depend either on this coordinate.
As shown in the literature \cite{born-wolf}, this allows us to express
the electric and magnetic fields as a sum of $E$-polarized ($E_{e,x} =
E_{e,y} = H_{e,z} = 0$) and $H$-polarized ($H_{h,x} = H_{h,y} = E_{h,z}
= 0$) components, i.e.,
\begin{equation}
 \begin{array}{l}
  {\bf E} = {\bf E}_e + {\bf H}_h
    = {\bf E}_e + \displaystyle \frac{i}{\omega\epsilon_0}
     \left[ \nabla \times {\bf H}_h \right] , \\
  {\bf H} = {\bf H}_e + {\bf H}_h
    = - \displaystyle \frac{i}{\omega\mu_0}
     \left[ \nabla \times {\bf E}_e \right] + {\bf H}_h .
 \end{array}
 \label{eq2}
\end{equation}

Let us assume that the incident EM field is a plane wave of frequency
$\omega$ and wave vector $k$ propagating along the $y$-axis, with
arbitrary polarization.
It can be written as \cite{sanz1}
\begin{equation}
 \begin{array}{l}
  \tilde{\bf E}_0 ({\bf r},t) = \left[
    - \beta e^{i(ky + \phi)} \hat{\bf x} + \alpha e^{iky} \hat{\bf z}
    \right] e^{-i\omega t} , \\
  \tilde{\bf H}_0 ({\bf r},t) =
    \displaystyle \sqrt{\frac{\epsilon_0}{\mu_0}} \left[
     \alpha e^{iky} \hat{\bf x} + \beta e^{i(ky + \phi)} \hat{\bf z}
    \right] e^{-i\omega t} ,
 \end{array}
 \label{eq3}
\end{equation}
where $\alpha$, $\beta$ and $\phi$ are real quantities.
When $\phi = 0$ or $\pi$ and $\alpha$ and $\beta$ are arbitrary, the
polarization of the incident wave is linear.
If $\phi = \pm \pi/2$ and $\alpha = \beta$, the polarization is
circular.
In all other cases the polarization is elliptic.
In (\ref{eq3}) we recognize the $E$-polarized and $H$-polarized
components,
\begin{equation}
 \begin{array}{l}
  {\bf E}_{0,e} = \alpha \Psi_0 \hat{\bf z} , \\
  {\bf H}_{0,h} = \beta \displaystyle \sqrt{\frac{\epsilon_0}{\mu_0}}
    \ e^{i\phi} \Psi_0 \hat{\bf z} ,
 \end{array}
 \label{eq4}
\end{equation}
with
\be
 \Psi_0 ({\bf r}) = e^{iky}
 \label{eq5}
\ee
being the corresponding scalar field satisfying Helmholtz's equation.

If the boundary conditions at the grating for $E_{e,z}({\bf r})$ and
$H_{h,z}({\bf r})$ are the same and both fields satisfy Helmholtz's
equation, we may also assume that these fields are proportional to a
scalar field, $\Psi({\bf r})$, which also satisfies the same boundary
conditions and the Helmholtz equation.
Accordingly, we can express these fields as
\begin{equation}
 \begin{array}{l}
  {\bf E}_e = \alpha \Psi \hat{\bf z} , \\
  {\bf H}_h = \beta \displaystyle \sqrt{\frac{\epsilon_0}{\mu_0}}
    \ e^{i\phi} \Psi \hat{\bf z} .
 \end{array}
 \label{eq6}
\end{equation}
Within the paraxial approximation and for completely transparent
slits with their support being totally absorbing, the solution to
Helmholtz's equation behind a double-slit reads as
\be
 \Psi(x,y) = \sum_{i=1,2} \psi_i(x,y)
  = \displaystyle \sqrt{\frac{k}{2\pi y}} \ e^{-i\pi/4} e^{iky}
   \sum_{i=1,2} \int_A \Psi_0(x',0^-) e^{ik(x-x')^2/2y} dx' ,
 \label{eq7}
\ee
where $\Psi_0(x',0^-)$ is the incident scalar wave just before the
grating.
Taking into account (\ref{eq2}), (\ref{eq6}) and (\ref{eq7}), we find
that
\be
 {\bf E} = {\bf E}_1 + {\bf E}_2 , \qquad
 {\bf H} = {\bf H}_1 + {\bf H}_2 ,
 \label{eq8}
\ee
where
\be
 \begin{array}{l}
  {\bf E}_i = \displaystyle
    \frac{i\beta e^{i\phi}}{k}
    \frac{\partial \psi_i}{\partial y} \ \hat{\bf x}
  - \frac{i\beta e^{i\phi}}{k}
    \frac{\partial \psi_i}{\partial x} \ \hat{\bf y}
  + \alpha \psi_i \hat{\bf z} , \\
  {\bf H}_i = \displaystyle
  - \frac{i\alpha}{\omega \mu_0}
    \frac{\partial \psi_i}{\partial y} \ \hat{\bf y}
  + \frac{i\alpha}{\omega \mu_0}
    \frac{\partial \psi_i}{\partial x} \ \hat{\bf y}
  + \frac{k \beta e^{i\phi}}{\omega \mu_0} \ \psi_i \hat{\bf z}
 \end{array}
 \label{eq9}
\ee
represent, respectively, the electric and magnetic fields propagating
from the $i$th slit, with $ = 1, 2$.

As seen above, experimentally one measures the light intensity, i.e.,
the EME density on the screen positioned at some distance from the
diffraction grating.
From the general expression for the EME energy density,
\be
 U({\bf r}) = \displaystyle \frac{1}{4}
  \left[ \epsilon_0 {\bf E}({\bf r}) \cdot {\bf E}^*({\bf r})
   + \mu_0 {\bf H}({\bf r}) \cdot {\bf H}^*({\bf r}) \right] ,
 \label{eq10}
\ee
we find that the EME density of the incident EM wave (\ref{eq3}) is
constant (i.e., independent of both $x$ and~$y$),
\be
 U_0({\bf r}) = \displaystyle \frac{\epsilon_0}{2}
   \left( \alpha^2 + \beta^2 \right) .
 \label{eq11}
\ee
However, as a result of the interaction between the EM field and the
grating, the EME density behind this grating becomes dependent on $x$
and $y$,
\be
 U({\bf r}) = \frac{\epsilon_0}{4k^2} \left( \alpha^2 + \beta^2 \right)
  \left[
    \left\arrowvert \frac{\partial \Psi}{\partial x} \right\arrowvert^2
  + \left\arrowvert \frac{\partial \Psi}{\partial y} \right\arrowvert^2
  + k^2 |\Psi|^2 \right] .
 \label{eq12}
\ee
The solution (\ref{eq7}) to the Helmholtz equation satisfies the
approximate relations
\be
 \left\arrowvert \frac{\partial \Psi}{\partial x} \right\arrowvert \ll
  \left\arrowvert \frac{\partial \Psi}{\partial y} \right\arrowvert ,
  \qquad
  \left\arrowvert \frac{\partial \Psi}{\partial y} \right\arrowvert
  \approx ik\Psi
 \label{eq13}
\ee
and, therefore, the EME density (\ref{eq12}) becomes proportional to
$|\Psi|^2$, i.e.,
\be
 U({\bf r}) = \displaystyle \frac{\epsilon_0}{2}
  \left( \alpha^2 + \beta^2 \right) |\Psi|^2 =
  \frac{\epsilon_0}{2} \left( \alpha^2 + \beta^2 \right)
  \left[ |\Psi_1|^2 + |\Psi_2|^2 + 2 {\rm Re} (\psi_1 \psi_2^*)
    \right] .
 \label{eq14}
\ee
As can be noticed from this expression, the $x$ and $y$ dependence of
the EME density can be determined through $|\Psi|^2$.
The polarization of the incident EM wave has no influence on the
dependence of the EME density on $x$ and $y$ and, therefore, the
interference pattern will be independent of the incident field
polarization.
This distribution is the same for linear, circular or elliptic
polarization, as seen in the experimental results displayed in
Figs.~\ref{fig2} and \ref{fig3}.
It is important to stress that this conclusion arises from assuming
that the EM field does not depend on the $z$-coordinate.

The independence of the EME density on polarization implies that the
interference pattern will be the same for both linearly polarized and
natural light, in agreement with one of the empirical findings observed
by Arago and Fresnel, namely the first Arago-Fresnel law.
This law states that ``two rays polarized in one and the same plane
act on or interfere with each other just as natural rays, so that the
phenomena of interference in the two species of light are absolutely
the same'' \cite{arago,barakat}.
The generalization of this law also includes elliptic polarization,
and circular polarization as a special case of elliptic polarization
\cite{mujat}.


\subsection{EM field and EME density behind a double-slit grating
followed by orthogonal polarizers}
\label{sec32}

The approach described above can also be applied to the study of the EM
field and EME density behind two slits which are followed by two linear
orthogonal polarizers.
Now, instead of considering Eqs.~(\ref{eq6}), the $E$-polarized and
$H$-polarized components of the EM field are expressed \cite{sanz1}
in terms of $\psi_1$ and $\psi_2$, as
\begin{equation}
 \begin{array}{l}
  {\bf E}_e = \alpha \psi_1 \hat{\bf z} , \\
  {\bf H}_h = \beta \displaystyle \sqrt{\frac{\epsilon_0}{\mu_0}}
    \ e^{i\phi} \psi_2 \hat{\bf z} .
 \end{array}
 \label{eq15}
\end{equation}
Thus, by substituting (\ref{eq15}) into (\ref{eq2}), we obtain the EM
field behind the two slits covered by the orthogonal polarizers,
\be
 \begin{array}{l}
  {\bf E}_i = \displaystyle
    \frac{i\beta e^{i\phi}}{k}
    \frac{\partial \psi_2}{\partial y} \ \hat{\bf x}
  - \frac{i\beta e^{i\phi}}{k}
    \frac{\partial \psi_2}{\partial x} \ \hat{\bf y}
  + \alpha \psi_1 \hat{\bf z} , \\
  {\bf H}_i = \displaystyle
  - \frac{i\alpha}{\omega \mu_0}
    \frac{\partial \psi_1}{\partial y} \ \hat{\bf y}
  + \frac{i\alpha}{\omega \mu_0}
    \frac{\partial \psi_1}{\partial x} \ \hat{\bf y}
  + \frac{k \beta e^{i\phi}}{\omega \mu_0} \ \psi_2 \hat{\bf z} .
 \end{array}
 \label{eq16}
\ee
From this EM field, the expression for the EME density will read as
\begin{eqnarray}
 U({\bf r}) & = & \frac{\epsilon_0}{2} \ \alpha^2 \left[
   \left\arrowvert \frac{\partial\psi_1}{\partial x} \right\arrowvert^2
 + \left\arrowvert \frac{\partial\psi_1}{\partial y} \right\arrowvert^2
 + k^2 |\psi_1|^2 \right]
 + \frac{\epsilon_0}{2} \ \beta^2 \left[
   \left\arrowvert \frac{\partial\psi_2}{\partial x} \right\arrowvert^2
 + \left\arrowvert \frac{\partial\psi_2}{\partial y} \right\arrowvert^2
 + k^2 |\psi_2|^2 \right] \nonumber \\
 & \approx & \frac{\epsilon_0}{2} \left( \alpha^2 + \beta^2 \right)
  \left[ |\psi_1|^2 + |\psi_2|^2 \right] .
 \label{eq17}
\end{eqnarray}
Under the presence of orthogonal polarizers behind the slits, the
EME density becomes the simple sum of EME densities, which spread
out independently from each slit.
As a consequence, interference fringes are absent.
This is in agreement with the second law experimentally established by
Arago and Fresnel \cite{arago}, namely the second Arago-Fresnel law
\cite{barakat}.
This law states that ``two rays primitively polarized in opposite
planes (i.e., at right angles to each other) have no appreciable action
on each other, in the very same circumstances where rays of natural
light would interfere so as to destroy each other''.
The generalization of this law also includes elliptic and circular
polarization of opposite handedness \cite{mujat}.


\section{EME-flow-line description }
\label{sec4}

\begin{figure}
 \bc
 \includegraphics[width=10cm]{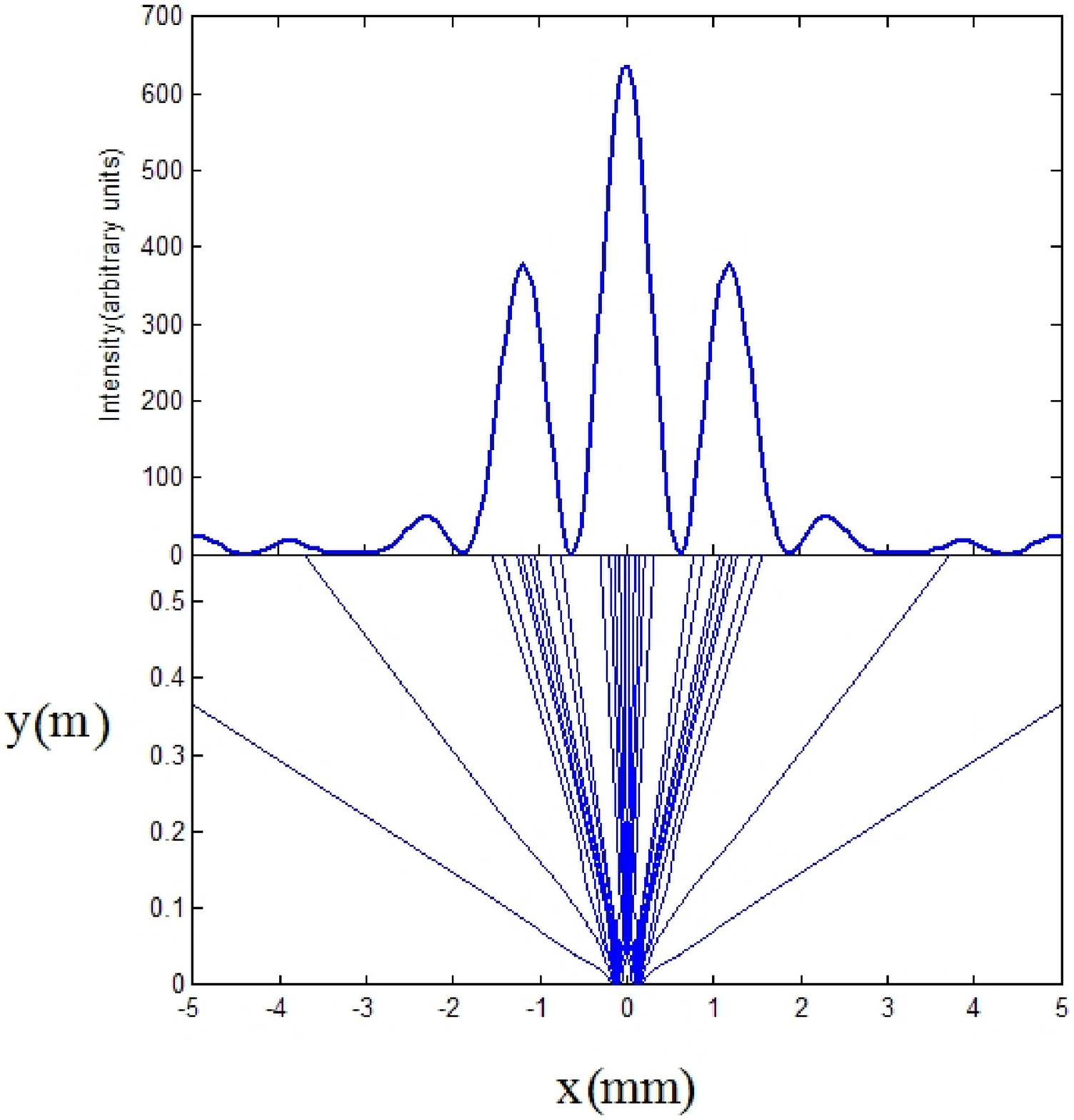}
 \includegraphics[width=12cm]{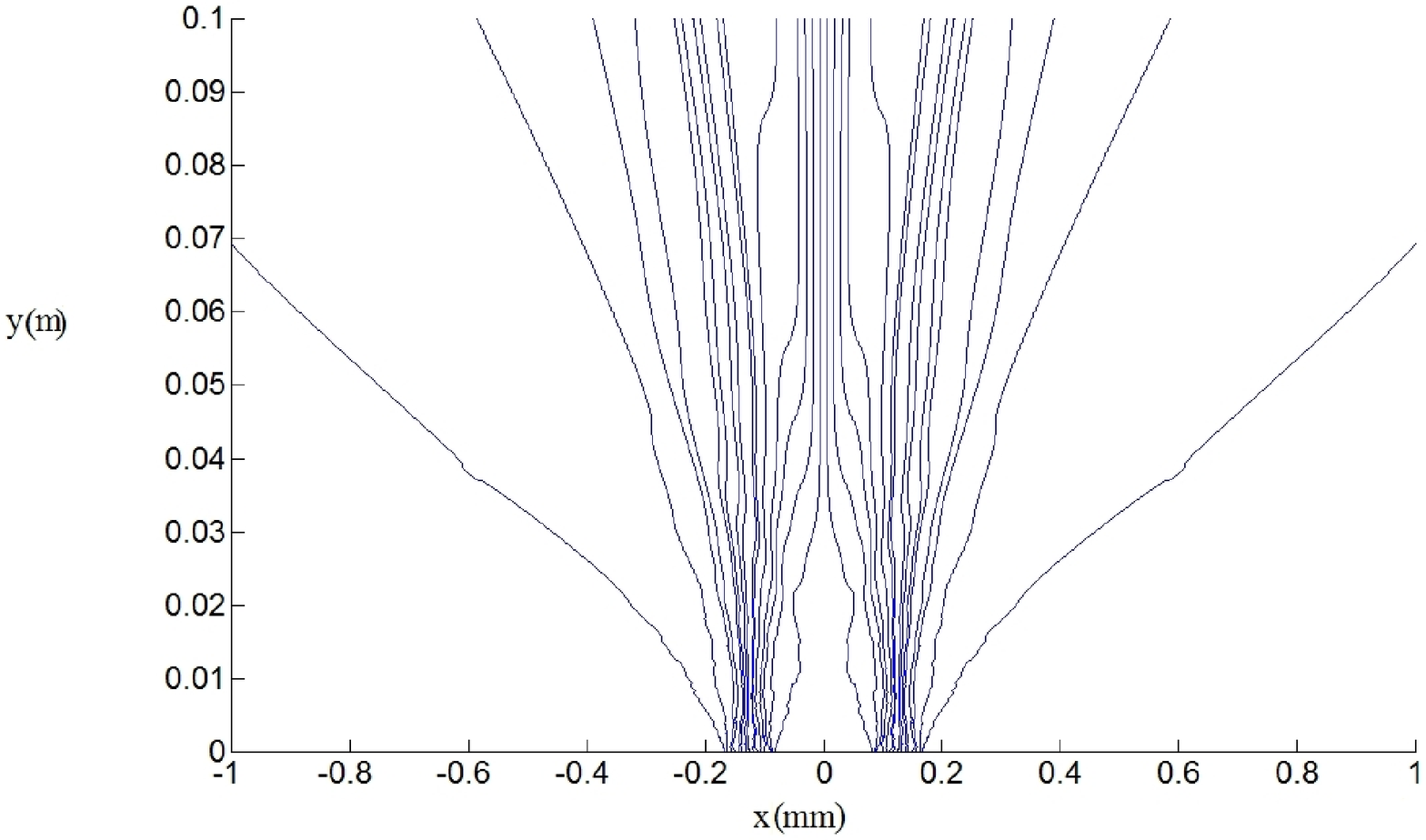}
 \ec
 \vspace{-4mm}
 \caption{\label{fig5} {\bf Top:} EME density at $L = 558$~mm from a
  double-slit grating illuminated by a circularly polarized laser beam
  of wavelength $\lambda = 532.5$~nm.
  {\bf Centre:} 30 EME flow lines behind a double-slit grating
  determined from (\ref{eq20}) and the scalar solution (\ref{eq7})
  of Helmholtz's equation.
  {\bf Bottom:} Enlargement of the EME flow lines in the near field
  region.
  The slits are assumed to be completely transparent, with their
  support being completely absorbing.
  The distance between the center of the slits is $d = 0.25$~mm and the
  slit width is $\delta = 0.1$~mm.}
\end{figure}

The properties of the EME density behind a grating and the generalized
Arago-Fresnel laws may be better understood with the aid of EME flow
lines, which can be interpreted as photon trajectories \cite{sanz1}.
The equation from which the EME flow lines are obtained is
\be
 \frac{d{\bf r}}{ds} = \displaystyle \frac{1}{c}
   \frac{{\bf S}({\bf r})}{U({\bf r})} ,
 \label{eq18}
\ee
where ${\bf S}({\bf r})$ is the real part of the complex Poynting
vector,
\be
 {\bf S}({\bf r}) = \displaystyle
  \frac{1}{2} \ {\rm Re} \left[ {\bf E}({\bf r}) \times
   {\bf H}^*({\bf r}) \right] ,
 \label{eq19}
\ee
$U({\bf r})$ is the time-averaged EME density and $s$ is the certain
arc-length of the corresponding path.
The substitution of Eqs.~(\ref{eq7})-(\ref{eq9}) into (\ref{eq19})
renders the trajectory equations along each direction,
\be
 \begin{array}{l}
  \displaystyle \frac{dx}{ds} =
    \frac{i\epsilon_0 (\alpha^2 + \beta^2)}{4kU({\bf r})}
    \left[ \Psi\ \frac{\partial \Psi^*}{\partial x}
         - \Psi^*\ \frac{\partial \Psi}{\partial x} \right] , \\[.4cm]
  \displaystyle \frac{dy}{ds} =
    \frac{i\epsilon_0 (\alpha^2 + \beta^2)}{4kU({\bf r})}
    \left[ \Psi\ \frac{\partial \Psi^*}{\partial y}
         - \Psi^*\ \frac{\partial \Psi}{\partial y} \right] , \\[.4cm]
  \displaystyle \frac{dz}{ds} =
  - \frac{i\epsilon_0\alpha\beta\sin\phi}{2k^2 U({\bf r})}
    \left[
    \frac{\partial \Psi}{\partial x} \frac{\partial \Psi^*}{\partial y}
  - \frac{\partial \Psi}{\partial y} \frac{\partial \Psi^*}{\partial x}
    \right] .
 \end{array}
 \label{eq20}
\ee
The projection of the EME flow lines on the $XY$-plane behind a
double-slit grating illuminated by circularly polarized laser light,
determined from Eqs.~(\ref{eq20}) and the corresponding experimental
conditions, are plotted in Fig.~\ref{fig5}.
As can be noticed, the EME flow lines (photon trajectories) accumulate
according to $U({\bf r})$ \cite{sanz1}, as given by Eq.~(\ref{eq14}).
It is important to stress that, in the case of circular and elliptic
polarization, EME flow lines do not remain within the $XY$-plane, but
there is a flow of energy parcels (photons) along the vertical axis.
Nevertheless, the projection of these three-dimensional trajectories
onto the $XY$-plane does not depend on the phase angle, $\phi$.
Due to this fact, the resulting
distribution of end points associated with the trajectories along
the x-axis does not depend on polarization. This is consistent
with the conclusions derived in the previous section from analysis
of the EME density dependence on polarization.

\begin{figure}
 \bc
 \includegraphics[width=10cm]{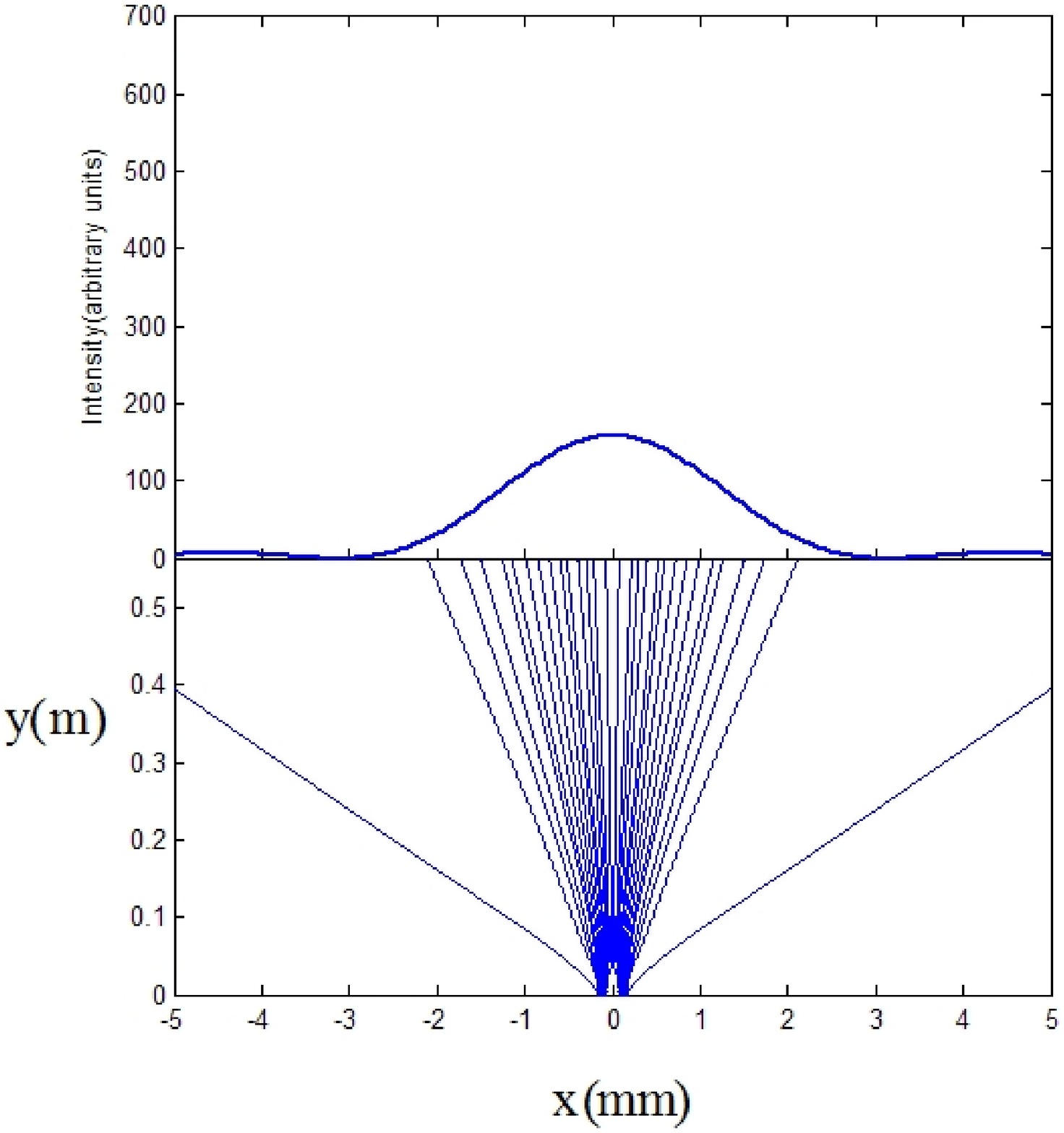}
 \includegraphics[width=12cm]{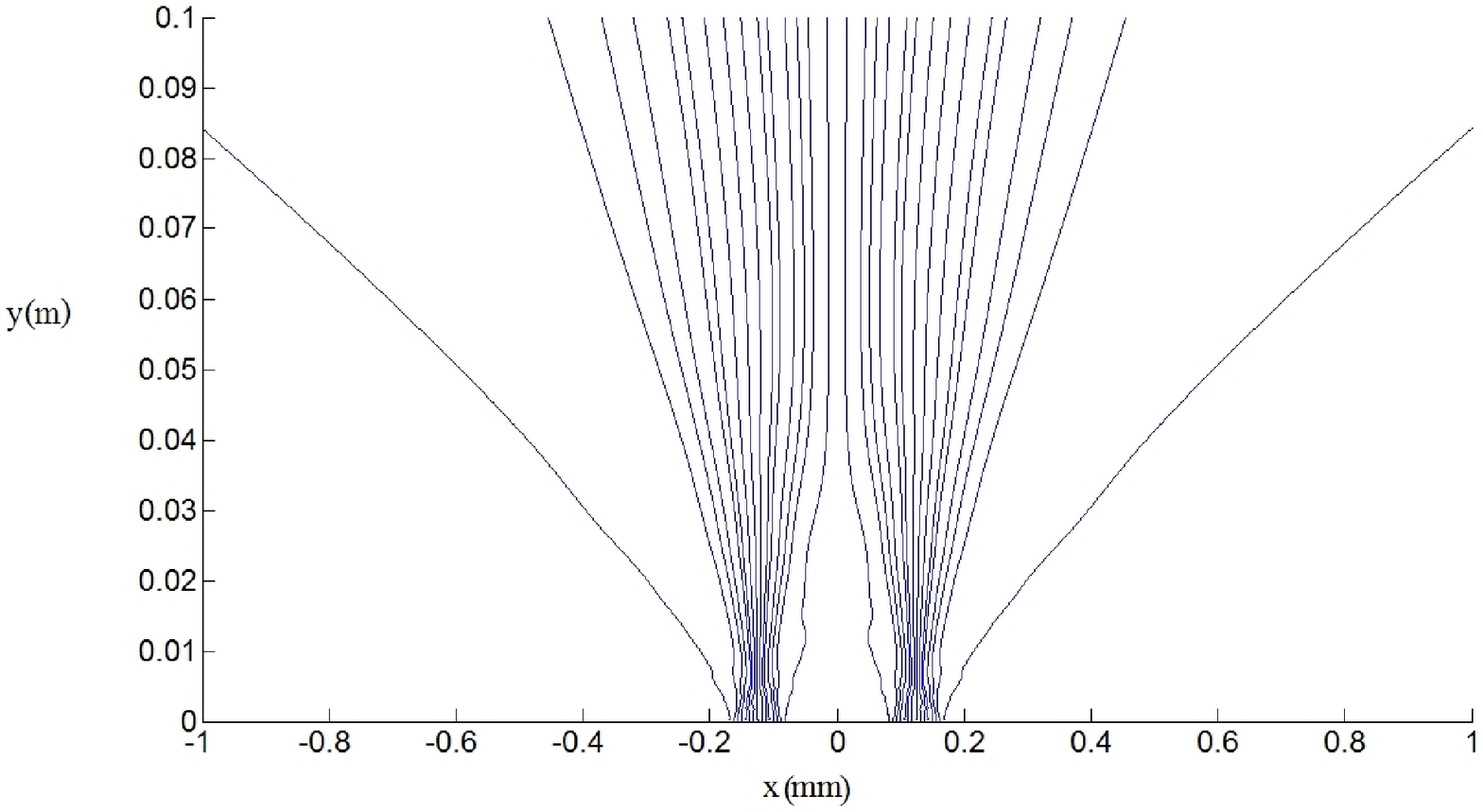}
 \ec
 \vspace{-4mm}
 \caption{\label{fig6} {\bf Top:} EME density at $L = 558$~mm from a
  double-slit grating with orthogonal polarizers upon the slits and
  illuminated by a circularly polarized laser beam of wavelength
  $\lambda = 532.5$~nm.
  {\bf Centre:} 30 EME flow lines behind a double-slit grating,
  determined from Eqs.~(\ref{eq16}), (\ref{eq18}) and (\ref{eq19}),
  and the scalar solution (\ref{eq7}) of Helmholtz's equation.
  {\bf Bottom:} Enlargement of the EME flow lines in the near field
  region.
  The slits are assumed to be completely transparent, with their
  support being completely absorbing.
  The distance between the center of the slits is $d = 0.25$~mm and the
  slit width is $\delta = 0.1$~mm.}
\end{figure}

The number of bright and dark fringes obtained in the experiment,
where $d = 0.25$~mm, $\delta = 0.1$~mm and $\lambda = 532.5$~nm, agrees
fairly well with the corresponding number predicted by the probability
amplitude of transverse momenta,
\be
 c(k_x) = \displaystyle \frac{1}{\sqrt{2\pi}}
  \int_{-\infty}^\infty dx' \Psi(x',0^+) e^{-ik_x x'}
  = \frac{1}{\sqrt{2\pi}} \sum_{i=1,2}
  \int_A dx' \Psi_0(x',0^-) e^{-ik_x x'} ,
 \label{eq21}
\ee
which, in the case of two slits, is given by
\be
 c(k_x) = \frac{2}{\sqrt{\pi\delta}}
  \frac{\sin(k_x\delta/2)}{k_x} \ \cos(k_x\delta/2) .
 \label{eq22}
\ee
The centers for the zeroth (central), first ($\pm 1$) and second
($\pm 2$) bright fringes are determined through the relation
$\cos(k_x\delta/2) = \pm 1$, which renders $k_{x,0} = 0$, $k_{x,\pm 1}
= \pm 2\pi/d$ and $k_{x,\pm 2} = \pm 4\pi/d$, respectively.
On the other hand, the centers of the dark fringes are obtained taking
into account that $\cos(k_x\delta/2) = 0$ and, therefore, $k_x = \pm
(2n+1) \pi/d$, with $n = 0, 1, 2, \ldots$.

The intensity of the bright fringes decreases because of the envelope
of the interference pattern, which is given by first factor and
describes the single-slit diffraction.
This factor vanishes as $k_x$ increases from zero to the value
determined by the relation $\sin(k_x\delta/2)$, i.e., $k_x = \pm
2\pi/\delta$.
The point where the intensity falls to zero due to diffraction will
coincide with one of the interference minima if
\be
 \frac{2\pi}{\delta} = \frac{(2n+1)\pi}{d} \quad \longrightarrow
  \quad \frac{d}{\delta} = \frac{2n+1}{d} .
 \label{eq23}
\ee
As seen in the experimental results, this condition holds, because
$d/\delta = 5/2 = 2.5$.
Therefore, the second bright fringe will be very weak (almost fainting)
and the third dark fringe will be negligible in comparison with the
remaining continuous dark pattern, as seen in Figs.~\ref{fig2} and
\ref{fig3}.

When mutually orthogonal polarizers are positioned upon the slits,
both the EME distribution and the EME flow lines change dramatically
\cite{sanz1}.
The resultant distribution of trajectory end points along the $x$-axis
will be consistent with expression (\ref{eq17}) for the EME density and
interference fringes will be absent.


\section{Summary and conclusions: EME flow lines versus ``which-slit''
information}
\label{sec5}

The analysis presented above throws some novel light on the first and
second Arago-Fresnel laws.
Moreover, it also results relevant to the discussion on the
wave-particle duality of light.
The fact that fringes disappear after covering the slits with
orthogonal polarizers has been used by the proponents of the principle
of complementarity to affirm that information about the path destroys
the interference \cite{scully,walborn}.
Within this type of argumentations or interpretations, ``path'' just
means the slit through which the photon passes and any discussion about
the whole path, from the grating to the screen, is thus forbidden.

Argumentations based on ``which-slit'' information are closely
related to the reasoning leading to the notion of quantum erasers
\cite{scully,walborn}.
By applying this reasoning to the sequence in which the first
experiment is a double-slit experiment with orthogonal polarizers and
the second experiment is a double-slit experiment without orthogonal
polarizers, one could assign the lack of interference to the existence
of ``which-slit'' information.
The appearance of interference after removing the polarizers could
be then attributed to erasing and, consequently, to the loss of
information about the slit which the energy parcel or photon passed
through.

The interpretation used here considers a set of entire paths from the
grating to the screen, such paths being determined from the EM field
and the Poynting vector.
EME flow lines starting from slit~1 will end up in the left-hand side
part of the screen, while those starting from slit~2 will end up in
the right-hand side, as also shown in quantum mechanics for matter
particles \cite{sanz2}.
This is valid both when interference is present and when no
interference fringes are observed \cite{sanz3,sanz4}.
But, the distribution of this EME flow lines is different in the two
cases.
In the absence of polarizers, the distribution shows interference
fringes (Fig.~\ref{fig5}); in the presence of polarizers, the fringes
are absent (Fig.~\ref{fig6}), in perfect agreement with experimental
results shown at Fig.~\ref{fig4}.
Hence, observer's information about the slit through which photon
(energy parcel) went through is not relevant for the existence of
interference.
What is relevant it is the form of the EME field and its corresponding
form, which will model consequently the distribution of trajectories
and their topology.


\section*{Acknowledgments}

This article is written in honor of Margarita Man'ko with occasion of
her 70th birthday.
Two of us, M.B.\ and M.D., have enjoyed for many years in scientific
contacts and collaboration with Margarita, her husband Vladimir Man'ko
and her daughter Olga Man'ko.
All authors appreciate very much the contribution of Margarita Man'ko
to the rise and development of the workshop series Central European
Workshop on Quantum Optics (CEWQO).
In particular, we appreciate her support for making that the 15th CEWQO
could take place in Belgrade in 2008.
This conference gave the authors the chance to meet and discuss
problems on quantum interference, and start the collaboration which
has given rise to the work presented here.

M.B.\ and M.D.\ acknowledge support from the Ministry of Science
of Serbia under Project ``Quantum and Optical Interferometry'',
No.~141003.
S.M.-A.\ and A.S.S.\ acknowledge support from the Ministerio de Ciencia
e Innovaci\'{o}n (Spain) under Project FIS2007-62006; A.S.S. also thanks
the Consejo Superior de Investigaciones Cient\'{\i}ficas for a JAE-Doc
Contract.
T.L.D.\ acknowledges financial support from the University of Fribourg,
which has allowed to carry out the experiments presented in this work.


\end{document}